\documentclass[aps,prl,showpacs,floats,twocolumn,floats,superscriptaddress,floatfix]{revtex4}
\usepackage{bm}

\usepackage{times}
\usepackage{verbatim}
\usepackage{graphicx}
\usepackage{graphics,epsfig}

\usepackage{theorem}
\usepackage{makeidx}
\usepackage{amsmath}
\usepackage{epic}
\usepackage{amscd}

\usepackage{bbm}
\usepackage{xy}
\usepackage{amssymb}
\usepackage{epsfig}

\begin{document}
\newif\iffigs 
\figstrue

\iffigs \fi

\def\drawing #1 #2 #3 {
\begin{center}
\setlength{\unitlength}{1mm}
\begin{picture}(#1,#2)(0,0)
\put(0,0){\framebox(#1,#2){#3}}
\end{picture}
\end{center} }

\newcommand{\ul}{{\bm u}_{\scriptscriptstyle \mathrm{L}}}
\newcommand{\deltal}{{\delta}_{\scriptscriptstyle \mathrm{L}}}

\newcommand{\ue}{\mathrm{e}}
\newcommand{\ui}{\mathrm{i}\,}
\newcommand{\kg}{{k_{\scriptscriptstyle \mathrm{G}}}}
\def\v{\bm v}
\def\x{\bm x}
\def\k{\bm k}
\def\ds{\displaystyle}

\title{Intermittency and Thermalization in Turbulence}

\author{Jian-Zhou Zhu}
\affiliation{T5 and CNLS, MS B258, Los Alamos National Laboratory, Los Alamos, NM 87545, U.S.A.}
\author{Mark Taylor}
\affiliation{{CCIM}, MS 0370, Sandia National Laboratories, Albuquerque, NM 87185, U.S.A.}

\begin{abstract}
A dissipation rate, which grows faster than any
power of the wave number in Fourier space, may be scaled to lead a hydrodynamic system
{\it actually} or {\it potentially} converge to its Galerkin truncation.  Actual convergence we name for the asymptotic truncation at a finite
wavenumber $k_G$ above which modes have no dynamics; and, we define potential convergence for the truncation at $k_G$ which, however, grows without bound. Both types of
convergence can be obtained with the dissipation rate
$\mu[cosh(k/k_c )-1]$ who behaves as $k^2$ (newtonian) and $\exp\{k/k_c\}$ for small and large $k/k_c$ respectively.  Competition physics of cascade, thermalization and dissipation are discussed with numerical Navier-Stokes turbulence, emphasizing on the intermittency growth.
\end{abstract}

\pacs{47.27.Gs; 05.20.Jj; 02.30.Jr}

\vspace{5pt}

\maketitle

Energy cascade in three-dimensional (3D) turbulence is accomplished by hierarchically generating smaller and weaker eddies (which are eventually eliminated by dissipation,) while thermalization in a conserved system is by sharing (generalized) energies among all excitations equally. Fermi et al. \cite{FPU} found some nonlinear systems did not simply thermalize as intuitively expected. However, it has been recently shown \cite{hgtbt} that  even some dissipative systems do partially thermalize, somewhat counterintuitively again. The idea is simply that the normal dissipation is just an incomplete truncation. The truncated Euler equations are a Liouville system and may thermalize to an equipartitioned $k^2$ spectrum in the three dimensional nonhelical case \cite{Lee52,Onsager}. It is then proposed to explain the flatter spectrum in between the inertial and dissipation ranges in fluid turbulence, usually called a bottleneck (see \cite{hgtbt} and references therein,) as partial thermalization. 

The notion of partial thermalization was proposed with the use of a high power $\alpha$ of the Laplacian (hyperviscosity) in the dissipative term of hydrodynamical equations. More precisely, consider the hydrodynamic $[\frac{ \partial }{ \partial t }+\mu (\frac{k}{k_G})^{2\alpha}]\hat{{\bf u}}({\bf k})=\hat{{\bf B}}({\bf u},{\bf u})$ for the velocity field ${\bf u}$ in a cyclic box represented in Fourier (wave number ${\bf k}$) space, with $k_G$ being off-lattice, and ${\bf B}(\cdot,\cdot)$ is the nonlinear term.
With $\alpha \to \infty$ while the positive but finite $\mu$ and $k_G$ being held fixed, the dynamics corresponds to the Galerkin-truncated-at-$k_G$ equation $\partial_t v = \Pi_{k_G} B(v,v)$: The truncation projector $\Pi_{k_G}$ is defined as a low-pass filtering operator which keeps the
harmonics with wavenumber less than $k_G$ and sets
the other ones to zero; $v=\Pi_{k_G} u$. 
We thus see that equilibrium statistical physics plays a role in competition with the unique dynamics controlled by the structure of dynamical euqations. On the other hand, inertial range of turbulence, characterized by a constant energy flux, is nonequilibrium, and it has been observed that some aspects of the intermittent cascade physics of turbulence also work beyond the inertial state (see, e.g., \cite{jzzPRE05}.) The interesting lesson we thus have learned is that turbulence is so generous as to allow the persistence of both these two extreme cases of statistical mechanics!

As, from the above, dissipation is responsible for both cascade and thermalization, what then is the essence of the dissipation mechanism needed to
obtain convergence of the dynamics to a Galerkin
truncation, and that thermalization? What if the asymptotic truncation wavenumber grows
without limit, in some particular way? And, how does thermalization affect the dyanamics
of the flow structures which are generally counted for turbulence intermittency?  Answers to them are essential to the general understanding of nonlinear dissipative systems and need relevant details of explicit examples. We start with a brief review discussion with our results:

{\it a)} We notice from Frisch et al. \cite{hgtbt} that the dissipation rate needs to grow faster than any power of $k$ to converge to a Galerkin-truncation operator of the system. Exponential growth, among others, of the dissipation rate will be shown to be also able to lead to such Galerkin truncation of a hydrodynamic system. Dissipation rate 
\begin{equation}\label{eq:cosh1}
\mathcal{D}(k)=\mu [\cosh(k/k_c)-1],
\end{equation} now called {\it coshcosity}, which grows exponentially for large $k$ is expected \cite{cosh} to be enough to tame the solutions to be not only analytic but also entire, is a perfect model for our purposes here. 

{\it b)} It was argued that the Galerkin-truncation for a hydrodynamic system has a mathematical consistency problem in the sense that the symplectic structure and Casimirs are broken, and a self-consistent sine truncation (for two-dimension hydrodynamics) is then proposed \cite{Zeitlin}. One should notice that with the presence of growing dissipation rate at high modes the Galerkin truncation is actually physically relevant \cite{hgtbt}. And, we do not even know what the Euler equations are \cite{Onsager, FrischBook} so that discussing convergence to 3D Euler equations seems pre-mature. However, we may numerically obtain some indication of what would happen when $k_G$ goes to infinity in some specific way. We denote one type of asymptotic behavior as ``potential convergence" to Galerkin truncation.

{\it c)} Boyd \cite{BoydJSC94} studied various dissipation models to observe their effects on a shock structure. The statistical consequences on turbulence should then also be investigated in detail. In turbulent flows, coherent flow structures \cite{SHEandCHEN} appear to be ``spotty" as the insight of intermittency given by Onsager \cite{Onsager}. We will present numerical Navier-Stokes results showing that small-scale coherence is destructed to be mistily and uniformly distributed with a reduction of intermittency. 

The local dissipativity $r(k)=\frac{1}{2}\frac{d \ln \mathcal{D}(k)}{d \ln k}$ in Eq. (\ref{eq:cosh1}) corresponds to $\alpha$ in hyperviscosity $\mathcal{D}(k)=\mu (\frac{k}{k_G})^{2\alpha}$. When $\alpha$ is increased without limit hyperviscous $\mathcal{D}(k)$ grows faster and faster, approaching zero below $k_G$ and infinitely large above $k_G$ (so that energy can not ``tunnel" through any more.) Now $r(k)$ from the coshcosity model grows with $k$ without limit, implying the potential of leading to the Galerkin truncation in a unique way. As the dissipation time scale is $\mu^{-1} [\cosh(k/k_c)-1]^{-1}$, with the eddy turn-over time fixed or varying slowly we can increase the dissipation wave number $k_d$ by decreasing $\mu$ with fixed $k_c$. This way, $k_d$ essentially becomes a truncation wavenumber $k_G$. Since $k_G$ is unlimited with $\mu \to 0$, we say {\it potential} convergence to distinguish from the {\it actual} convergence with finite $k_G$ as in \cite{hgtbt} and in the other parameterization of coshcosity immediately below. Similar results may be obtained by letting $\alpha$ be some increasing function of $k_G$ in the hyperviscosity model. Actually results below seem to indicate that, if $r(k)$ grows with $k$ fast enough, a limiting system with conservative dynamics may be obtained.

We now study how to obtain actual convergence to the Galerkin truncation at some finite wave number $k_G$ with the coshcosity model. Following the idea of the estimation of the dissipation wave number $k_d$ in the above, we should have the cross wave number $k_c$ decrease with $\mu$ to prevent $k_d$ from growing but to make it approach some finite value. This may be accomplished with the following observation: The right hand side of Eq. (\ref{eq:cosh1}) is essentially $\mu e^{k/k_c}/2$ for large $k$, so by letting $k_c=-k_G/\ln \mu$, as $\mu \to 0^+$, 
\begin{equation}\label{eq:cosh2}
\mathcal{D}(k) \sim \mu^{1-k/k_G}.
\end{equation} $\mathcal{D}(k)$ then goes to zero for $k<k_G$ and to infinity for $k>k_G$, which may lead to $actual$ convergence to the Galerkin truncation. Noticing that when $k$ is small, the coshcosity reduces to normal viscosity by taking the leading order term in the Taylor expansion, for $\mu$ not small enough we need to make some subtle adjustments, e.g., to take $k_c=k_G/\ln(\kappa+\sqrt{\kappa^2-1})$ with $\kappa=1/\mu+1$ as used below, to make the peak of the energy spectrum approach $k_G$ in a consistent way.

\begin{figure}[h!]
\begin{center}
\includegraphics[width=\columnwidth]{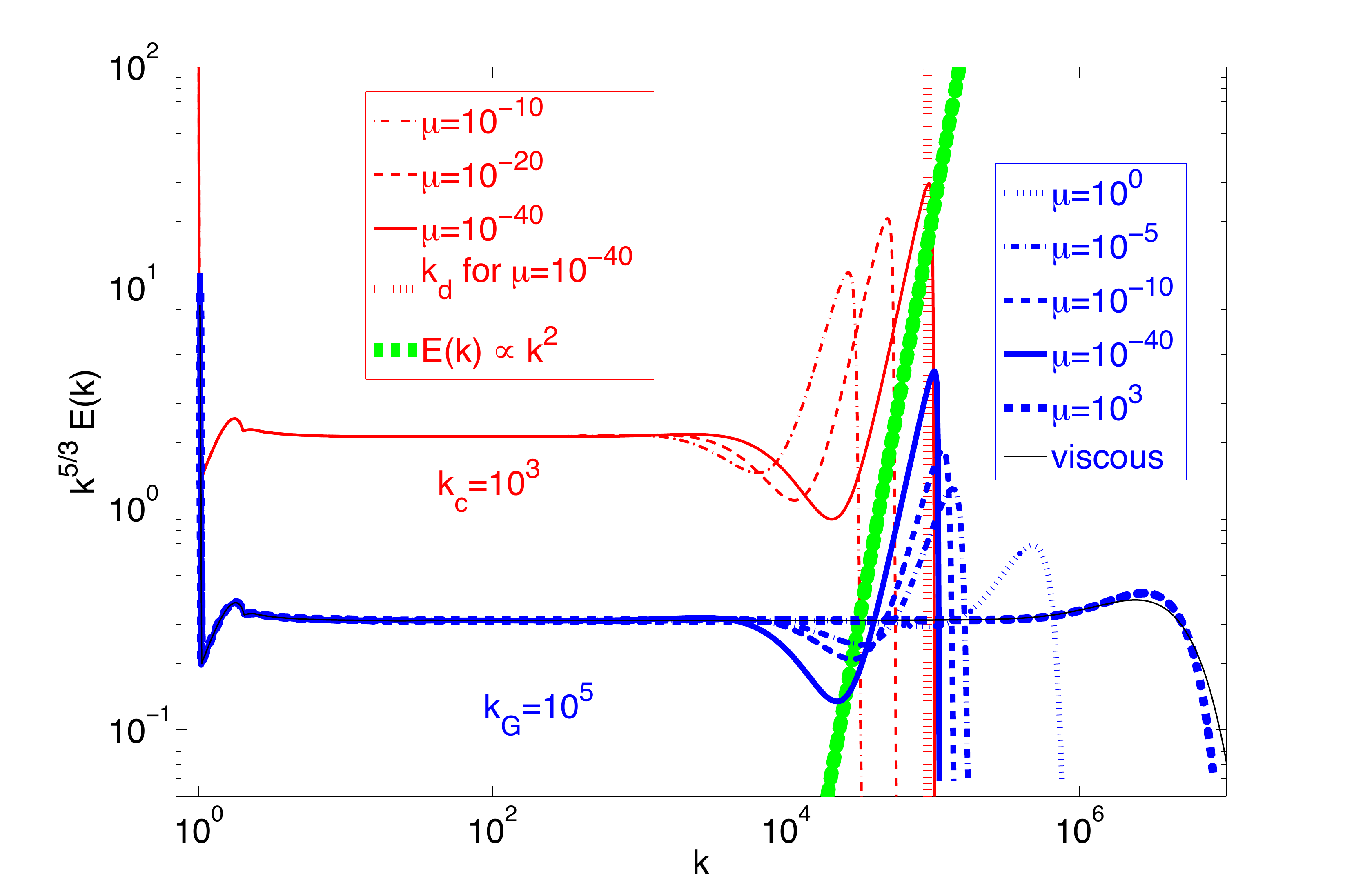}
\caption{EDQNM spectra with coshcosity (\ref{eq:cosh1}): $k_c=k_G/\ln(\kappa+\sqrt{\kappa^2-1})$ and $\kappa=1/\mu+1$ (lower blues lines), and $k_c=constant$ (upper red lines.) Fixed values of $k_G$ and $k_c$ are $10^5$ and $10^3$ respectively. The line named {\it viscous} is the same spectrum in \cite{hgtbt} with hyperviscosity $\mathcal{D}(k)=\mu (\frac{k}{k_G})^{2\alpha}$ for $\alpha=1$ case (actually normal viscosity now) with $\mu=1$ and $k_G=10^5$.}\label{fig:CoshApP}
\end{center}
\end{figure}

To further illustrate the ideas in the above we first perform the integrations of eddy-damped quasi-normal Markovian (EDQNM) equation for turbulence energy spectra as in \cite{hgtbt}. The equations read
\cite{EDQNM}
\begin{equation}
\left\{
\begin{array}{l}
\ds \left[\partial_t +2\mathcal{D}(k)\right]
E(k,t)  
=
\int\!\!\!\!\int_{\triangle_k}dpdq\,\theta_{kpq}\,\times \\[1.6ex]
\ds b(k,p,q)\frac{k}{pq}E(q,t)\left[k^2E(p,t)-p^2E(k,t)\right]+F(k)\;,\\[1.6ex]
\ds b(k,p,q)=\frac{p}{k}(xy+z^3)\;, \theta_{kpq}=\frac{1}
{\mu_k+\mu_p+\mu_q}\;, \\[1.6ex]
\ds \mu_k = \mathcal{D}(k) +\lambda
\left[\int_0^k p^2E(p,t)dp\right]^{\frac{1}{2}}\;.
\end{array}
\right.
\label{EDQNM}
\nonumber
\end{equation}
Here, $E(k,t)$ is the energy spectrum at time $t$, $F(k)$ is the
(spatial) forcing spectrum,  $x$, $y$, $z$ are the cosines of the
angles of the triangle with sides $k$, $p$, $q$ and $\lambda$ is a constant
related to the Kolmogorov constant.
We use the same forcing, discretization, and, mixed time-marching and iteration schemes, except that the hyperviscosity is now replaced by coshcosity. Two types of parameterizations introduced above are applied resulting in
Fig. \ref{fig:CoshApP}
 where the stationary spectra show respectively, as $\mu \to 0$, {\it actual} convergence to Galerkin-truncation at a finite wave number $k_G$ (upper red lines) and {\it potential} convergence to truncation at $-k_c\ln \mu$ (lower blue lines.)
We see clearly a $k^{-5/3}$ inertial range followed by a little bit of secondary bottleneck (a largest overshoot of $3\%$ for $\mu=10^{-40}$ also can be seen at around $k=2000$ if zoomed in,) a pseudo-dissipation range, a thermalization range and finally a dissipation range. The traditional assumption of inertial scaling going straight down to dissipation scale for estimating dissipation scale is vitiated by partial thermalization. When the thermalization is strong, the dissipation rate changes drastically around $k_d$ which can be conveniently estimated numerically by taking $\mathcal{D}(k_d) \sim 1$, as long as the eddy turnover time is not extremely far from order one (which is generally the case in numerical simulations) and then can be captured by $\mathcal{D}(k_d+\delta k)$ with $|\delta k|$ being relatively small.
If $k_c$ is kept constant and only $\mu$ itself varies, following the above phenomenology, we have $k_d \sim -k_c \ln \mu$ as designated by the vertical dotted line ($\approx 92103$) for $\mu=10^{-40}$ case. The system has a dissipation wavenumber $k_d$ which goes to infinity as $\mu \to 0$ and become essentially $k_G$. One might conclude from the upper lines for potential convergence that a (``directional") limit of conservative spectral dynamics may be obtained in this way \footnote{Such numerical ``conclusion" is of course very weak; and, no clear relevance to weak limit of dynamic solutions can be seen.}.

Knowledge of how the thermalization and dissipation depend on the parameters may be useful for turbulence simulations and modeling. With $\mu$ large, the majority of the spectrum falls in the regime where coshcosity reduces to normal viscosity. Here we compare the normal viscosity case, exactly the same as the $\alpha=1$ case in \cite{hgtbt}, and a coshcosity case with $\mu=10^3$. The latter may be of practical interest: The beginning of the dissipation range and below is mostly as with a normal viscosity, to which the coshcosity is reduced for small $k$. Only at higher wavenumbers the faster growth of the dissipation rate is felt, with the fluctuations there being low and of little practical significance. Such case may be used to avoid wasting resolution without developing a serious bottleneck as also proposed in \cite{cosh}.

Having the concepts of actual and potential convergence to Galerkin truncation in mind, together with the knowledge of EDQNM spectral dynamics, we then proceed to integrate the 3D Navier-Stokes dynamics (with the nonlinear term $B(\cdot,\cdot)=-\check{i}k_m P_{ij} \sum_{{\bf p}+{\bf q}={\bf k}} \hat{u}_j({\bf p})\hat{u}_m({\bf q})$, $\check{i}^2=-1$ and the transverse projection operator $P_{ij}=(\delta_{ij}k^2-k_ik_j)/k^2$.) We explore the physical contents of cascade, thermalization and dissipation concerning the statistics of velocity increment $\Delta_r u=u(x+r)-u(x)$ which is a deep quantity \cite{Onsager, FrischBook}.
It suffices for our main purposes here to report the results of the smallest $\mu$ ($=10^{-40}$) case from a $512^3$ simulation in a cyclic box of period $2\pi$ with $k_G=120.67$. 

\begin{figure}[h!]
\begin{center}
\includegraphics[width=\columnwidth]{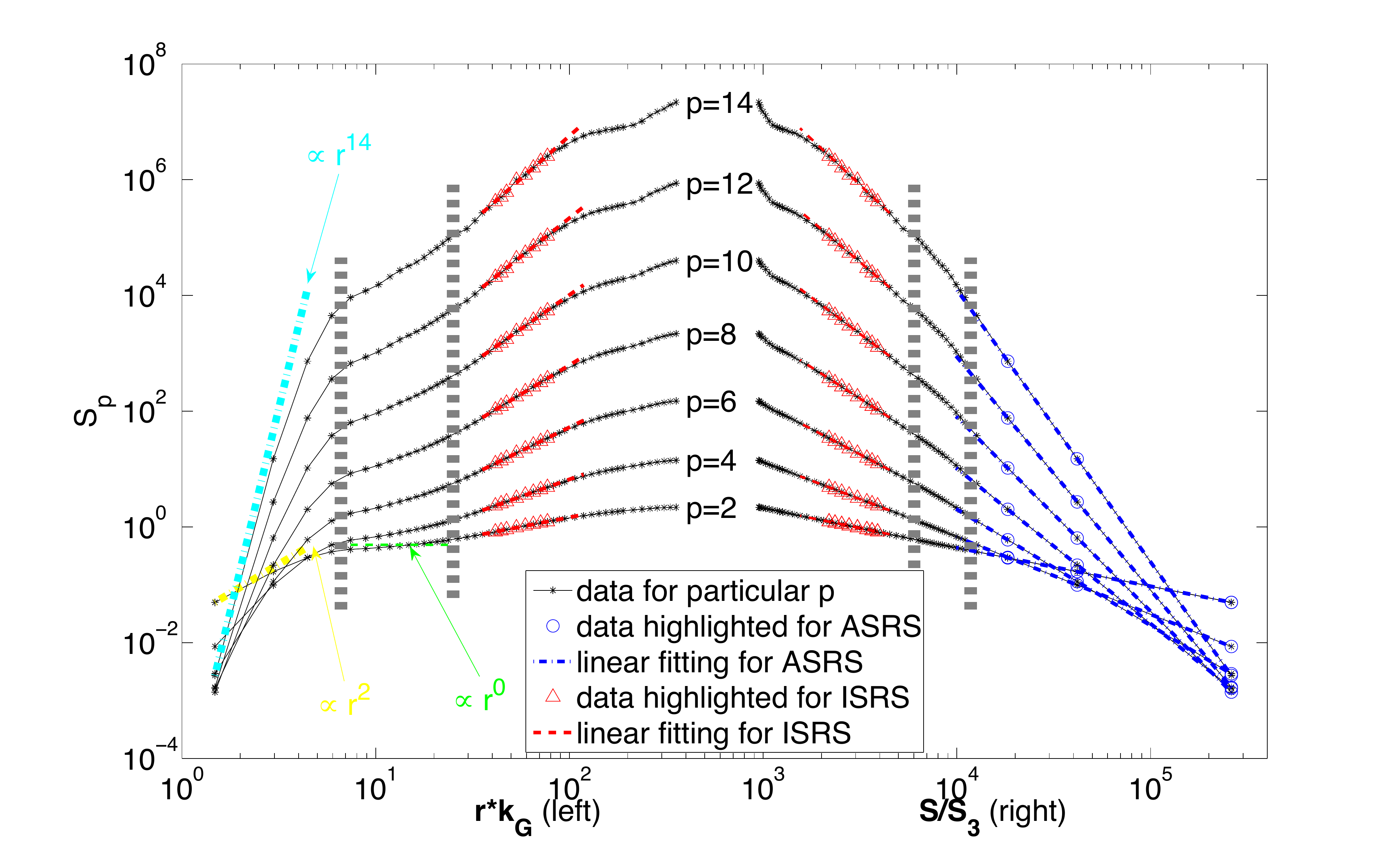}
\caption{Structure functions: Bottleneck scales are roughly designated by two vertical dashed lines. The analytical scalings for $p=2$ and $14$ and the scale-independent scaling ($\propto r^0$) are also plotted for reference. $S$ is a number used to shift the lines properly for the mirrored log-log plot against $S_3$.}\label{fig:CoshEm40Sp}
\end{center}
\end{figure}

Fig. \ref{fig:CoshEm40Sp} 
shows the structure functions $S_p(r)=<|\Delta_r u|^p>$ (with $<\bullet>$ meaning statistical average) against scales (left)  and against the third order structure function (right). We see that bottleneck scales in physical space now also clearly show up as designated by the corresponding pairs of vertical dashed lines. The fast damping beyond a scale $\eta$ in physical space can be related to the spacial discretization with grid width $\eta$. Identical independent Gaussian distribution, as of the equilibrium of Galerkin-truncated system in Fourier space, leads to also Gaussian distribution with negligible (decreasing with total number of grid points as can be shown with simple algebra) non-diagonal elements in the covariance matrix of the spacially discretized velocity (c.f. the $r^0$ reference scaling on the left plot.)  Clearly from the right part of the plot, $S_p$ against $S_3$, for the extended self-similarity (ESS - claiming power-law relations between structure functions $S_p(r) \sim [S_q(r)]^{\xi_{p,q}}$ \cite{ESS},) the bottleneck regime is a transitional range from analytical subrange scaling (ASRS, with the data highlighted with blue circles,) $\zeta^{A}_p=p$ to the inertial subrange scaling (ISRS) $\zeta^{I}_p$, with the data highlighted with red triangles: Although, from the left part of the figure we see that the pure analytical range is barely resolved ({\it cf.} the analytical scalings shown in the figure for $p=2$ and $14$ for comparison \footnote{For general purposes one does not need to resolve the extremely far dissipation range; see, e.g., T. Watanabe and T. Gotoh, J. Fluid. Mech. {\bf 590} 117 (2007).}) and a pure inertial range scarcely emerges, the right part of the figure shows that ESS works well \footnote{Discussions of the mathematical ``trivialness" of ESS can be found in D. Segel, V. L'vov and I. Procaccia, Phys. Rev. Lett., {\bf 76}, 1828 (1996) for a passive scalar model and in U. Frisch, S. Chakraborty and S. S. Ray, private communication (2009) for Burgers equation.} both for analytical scaling and inertial scaling but fails in the bottleneck scales - bulges can be found here for various $p$ with careful observation.
Using ESS, the analytical scaling is measured, through $S_p=A_pS_1^{\zeta_p^A}$, to be $\zeta_p^A=1.0076p$. The inertial scaling exponents $\zeta_p^I$ show no evidence of deviation from those measured in normal fluids. So, in our simulation, some asymptotic statistics in both inertial and analytical ranges are already available. 

\begin{figure}[h!]
\begin{center}
\includegraphics[width=\columnwidth]{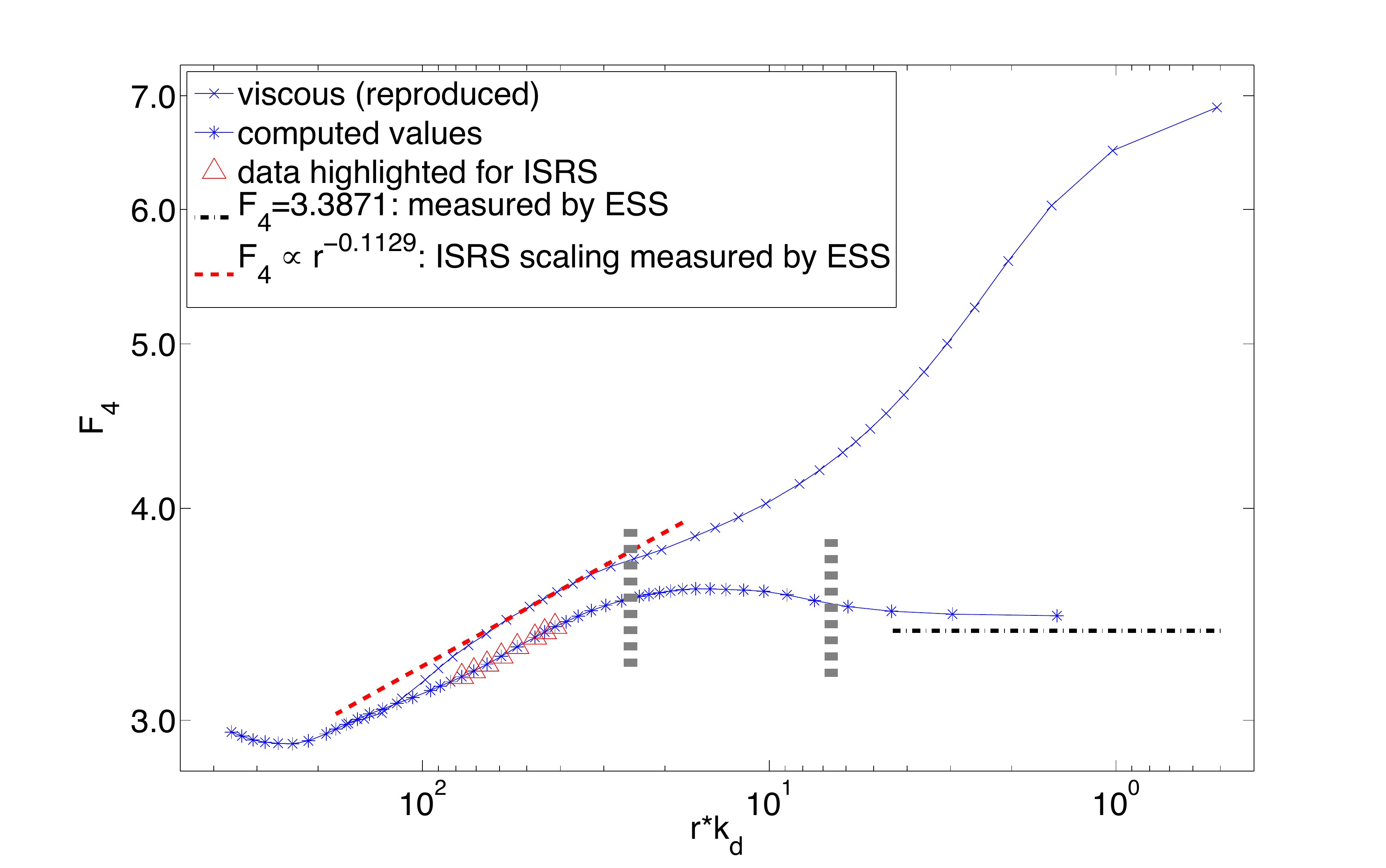}
\caption{Flatness factor for the velocity increments: The same bottleneck regime as in Fig. \ref{fig:CoshEm40Sp} is designated by a pair of dashed lines. A surrogate $A_4/A_2^2=3.3871$, by ESS, is measured for the asymptotic flatness factor in the analytical range $S_4/S_2^2=a_4/a_2^2$. The flatness inertial-range scaling (measured by ESS), with exponent $\zeta_4-2\zeta_2=-0.1129$, is also shown to fit well to the data of Navier-Stokes with normal viscosity (``viscous" - reproduced from \cite{zjzCPL06}).}\label{fig:rFlatness}
\end{center}
\end{figure}

Incompatibility of ESS in the bottleneck region has actually already been found in Ref. \cite{zjzCPL06} with the careful observation of the intermittency growth. As usual we thus measure the intermittency by the flatness factor of velocity increments $F_4(r)=S_4(r)/[S_2(r)]^2$. As shown in Fig. \ref{fig:rFlatness}
an obvious reduction of intermittency does present in the bottleneck regime, in comparison with the tiny  ``lull" with normal viscosity ({\it cf.} the ``viscous" data reproduced from \cite{zjzCPL06}.) Coherent vortex motions as a signification of the self-organization of the flow are randomized at such scales. Once $r$ goes into the analytical range with $S_p(r)=a_p r^p$, the flatness factor should be a constant depending on the settings of the flow, say, the Reynolds number (see, e.g., \cite{SSYnjp07}). The much smaller flatness factors in the far dissipation range for our coshcosity data than those with normal dissipation as shown in Fig. \ref{fig:rFlatness} may be further explained as follows. Strong damping of high modes push the complex singularities (if any) further away from the real axis(es) (even to infinity as proposed in \cite{cosh}) and/or weakening them. This then reduces the intermittency by the Frisch-Morf argument for the origin of intermittency \cite{FMpra81}. 

\begin{figure}[h!]
\begin{center}
\includegraphics[width=\columnwidth]{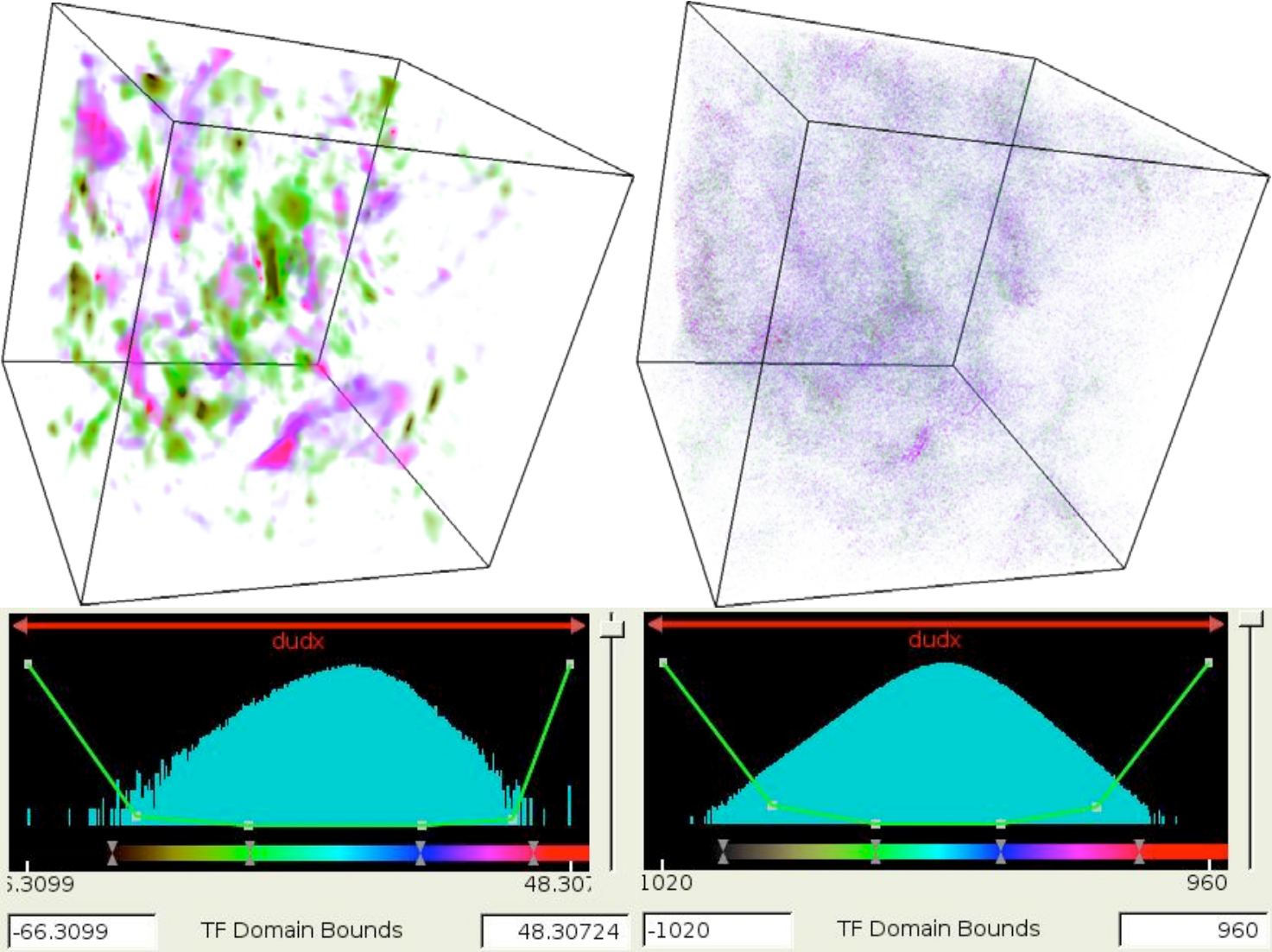}
\caption{Direct volume rendering \cite{vapor} of $\partial u/\partial x$: 
The top-left panel is the filtered field keeping only the modes under the thermalization range; the top-right panel is the total field; the lower panels present the linear-log plots of the histograms (of the corresponding fields of the upper panels) and the transfer functions. The $\partial u/\partial x$ values where the transfer functions (graphed by the lines connecting the opacity-control points)  for opacity beginning to be nonzero (for visibility) is approximately the values where contributions to flatness factors peak. 
Changing the transfer function will make the picture look more {\it clear} or {\it cloudy} instead of the {\it cleanly spotty} versus {\it mistily uniform} properties in the two renderings.
}\label{fig:Em40dudxDVR}
\end{center}
\end{figure}

It is then intriguing to see the effects on the flow configurations of the competition between thermalization and self-organization. The top-left panel of Fig. \ref{fig:Em40dudxDVR} 
shows the larger-scale structures are ```spotty' distribution of regions in which the velocity varies rapidly between neighboring points" as understood by Onsager \cite{Onsager}. Because, due to the eddy viscosity caused by strong thermalization of high modes which are filtered out now, lower modes below the thermalized range work like a normal fluid \cite{CichowlasPRL05}. These structures are embedded in the very-small-scale dissipative structures which are almost {\it mistily uniform} as shown in the top-right panel for the total field (including those thermalized modes.) 

We thus have shown by applications of the coshcosity model that a dissipation rate growing faster than any power may be parameterized to be of either potential or actual Galerkin truncation convergence capabilities. The dissipation wave number $k_d$ in either case approaches to the Galerkin-truncation wave number $k_G$ which grows without bound and is a finite value respectively. The EDQNM calculations for the latter case may indicate a limit of conservative spectral dynamics. Measurements and analyses of intermittency growth, ESS properties and careful visualization of the flow fields from a direct numerical simulation of Navier-Stokes equations detail the following understanding: Due to dissipation at small scales, turbulent flows are composed of competitive and complementary cascade and thermalization, which is pictured as self-organization and destruction of coherent structures at successive scales. 

This work was motivated and helped by communications with U. Frisch and A. Wirth, and, was supported by the CNLS LDRD program and the DOE ASCR Program in Applied Mathematics Research. We are grateful to the helpful discussions with J. Clyne, R. Ecke, G. Hammett, J. Herring, S. Kurien, T. Matsumoto, W. Pauls and V. Yakhot.

\end{document}